\documentclass[twocolumn,showpacs,preprintnumbers,amsmath,amssymb]{revtex4}
\usepackage{graphicx}% Include figure files
\usepackage{dcolumn}% Align table columns on decimal point
\usepackage{bm}% bold math
\usepackage{amssymb}%needed to use $\lesssim$ symbol

\begin{document}
\newcommand{\hesan}{$^3$He }
\newcommand{\heyon}{$^4$He }
\newcommand{\kake}{$\times$}

\title{Oscillation Frequency Dependence of Non-Classical Rotation Inertia
of Solid \heyon} \author{Y. Aoki} \author{J. C. Graves}
\author{H. Kojima}\affiliation{Serin Physics Laboratory, Rutgers University,
Piscataway, NJ 08854 USA}

%\date{\today}
\begin{abstract}
%The rotational inertia of the identical cylindrical solid $^4$He
%below 300 mK is studied at 496 and 1173 Hz by a double resonance
%torsional oscillator. At temperatures below 35 mK, the observed
%non-classical rotational inertia fraction at sufficiently low rim
%velocities is the same and the dissipation due to solid \heyon is
%close to zero for both modes.  Above 35 mK, the fraction is
%greater for the higher than the lower mode. The dissipation peak
%of the lower mode is greater than that of the higher mode by
%factor $\sim$ 1.7. Comparison of the drive dependence of the two
%modes shows that the observed reduction of the fraction is best
%characterized by critical velocity, \textit{not} amplitude nor
%acceleration, effects.  The critical velocity effects are shown
%to depend on the temporal history of how the rotational state of
%solid \heyon is reached.
The non-classical rotational inertia fraction of the identical
cylindrical solid \heyon below 300 mK is studied at 496 and 1173
Hz by a double resonance torsional oscillator. Below 35 mK, the
fractions are the same at sufficiently low rim velocities.  Above
35 mK, the fraction is greater for the higher than the lower
mode. The dissipation peak of the lower mode occurs at a
temperature $\sim$ 4 mK lower than that of the higher mode. The
drive dependence of the two modes shows that the reduction of the
fraction is characterized by critical velocity, \textit{not}
amplitude nor acceleration.
\end{abstract}

% \keywords{}

\maketitle

%\section{introduction}
%\subsection{general background motivation}
Ever since the observation of "Non-Classical Rotation Inertia"
(NCRI) of solid \heyon \cite{Leggett70} contained in their
torsional oscillators by Kim and Chan \cite{Kim04a,Kim04b},
apparent superfluidity in solid \heyon has motivated numerous
experimental and theoretical studies. A clear understanding of
the observed apparent partial decoupling between solid \heyon and
the container walls of torsional oscillators is lacking at
present. A number of interpretations of the observed decoupling
has been introduced including glassy state \cite{Nussinov06},
vortex liquid state \cite{Anderson07}, superglass
\cite{Boninsegni06}, and a phenomenological two component model
\cite{Huse07} (see \cite{Prokofev06} for a recent review). Many
of these theoretical ideas contain predictions on frequency
dependence of the magnitude of decoupling.  Although the
frequency of previous torsional oscillators has been varied
between 180 and 1500 Hz, the oscillators contained different
independently grown solid \heyon samples in different cells and
laboratories.  Here, we report on our measurements of NCRI of the
\emph{identical} solid \heyon sample contained in a torsional
oscillator having two resonant frequencies differing by a factor
of 2.4. We observe that the frequency dependence is a function of
both temperature ($T$) and the imposed oscillation velocity.
However, at sufficiently low temperature and oscillation
velocity, the measured superfluid fraction does not depend on
frequency.  At $T$ $>$ 35 mK, the superfluid fraction depends on
frequency, oscillation velocity and temperature.  In addition to
these frequency dependent effects, we show clear evidence for
hysteretic behavior of NCRI phenomenon depending on the history
of oscillation velocity and thermal processing from the normal
state above 300 mK to low temperature. Our observations put
severe constraints on some of the theoretical notions on the
interpretation of NCRI.

\begin{figure}[htbp]
\includegraphics[width=3in]{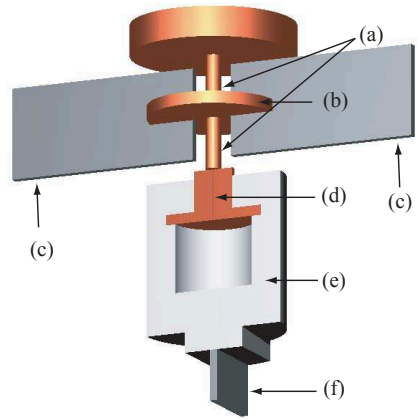}
\caption{\label{cell} Double resonance torsional oscillator. It
contains two masses and BeCu torsion rods (a) (2 mm outer
diameter). The upper mass includes a BeCu disc (b) with movable
detector and generator electrode fins (c). The lower mass is made
of a Stycast 1266 epoxy block (e) containing a hollow cylindrical
solid \heyon sample space (10.2 mm in diameter and 7.6 mm in
height. A second movable detector electrode (f) attached to the
bottom side of the lower mass is used to calibrate the upper
electrode detector motion. \heyon is filled through a central 0.8
mm diameter hole (d) drilled in the torsion rods and the upper
mass.  Fixed electrodes placed facing the movable electrodes are
not shown.}
\end{figure}

% description of cell
The conventional torsional oscillator having one resonant
frequency is made of one solid \heyon container attached to a
single torsion rod. Our torsional oscillator as shown in Fig.
\ref{cell} contains two masses, the upper "dummy" without solid
\heyon and the lower containing sample $^4$He, and two torsion
rods. The torsion rods, a portion of the dummy mass and the lower
flange fitted to the sample container are all machined from a
single stock of BeCu rod.  The first resonant mode (frequency
$f_1$ = 495.8 Hz) occurs when the two masses move in phase and
the second mode ($f_2$ = 1172.8 Hz) when the two masses move
180$^{\circ}$ out of phase. The phase relation of the modes are
verified by measuring the oscillator response at both the upper
and the lower detectors. Measurements of the oscillator response
are made with the upper electrode fins. The lower electrode is
used to calibrate oscillation amplitude at the sample cell
against the upper detector response. Owing to the uncertainty in
the capacitance between the lower fixed and movable electrodes,
the absolute oscillator amplitude is accurate only to within
$\pm$10 \%.  The relative accuracy in comparing the oscillator
amplitudes between the two modes is $\pm$5 \%.  The temperature
at the mixing chamber is measured by a calibrated \hesan melting
pressure thermometer. The temperature of the copper isolator
block which supports the oscillator is monitored by a carbon
resistance thermometer.

% characteristics of oscillator
The "background" characteristics \cite{Aoki07} of the resonant
frequencies, $f_{i0}$ ($i$ = 1, 2), and the quality factors
$Q_{i0}$ of our torsional oscillator with the sample cell empty
were measured as functions of temperature and drive level. The
$Q_{i0}$ ($1.4\times10^6$ and $0.63\times10^6$ for $i$ = 1 and 2,
respectively, at 30 mK) was determined from the ring down time.
The oscillation amplitude was calibrated against the measured
$Q_{i0}$ by the ring down time. The temperature dependence of the
resonant frequency and the quality factor of the empty cell were
both slightly dependent on the drive level but qualitatively
similar to those observed by others.  After the sample cell is
loaded with solid \heyon at 37 bar, for example, the resonant
frequencies measured at 300 mK decreased from the empty cell
values by $\Delta f_1$ = 0.667 Hz and $\Delta f_2$ = 2.049 Hz for
the first and second mode, respectively.

% description of sample preparation
Our sample solid \heyon is grown from commercially available
helium gas with nominal \hesan concentration of 0.5 ppm by the
blocked capillary method.  The cell is loaded near 3 K to a
desired pressure.  The part of the fill line which is thermally
anchored to the 1 K pot is quickly plugged with  solid while
\heyon in the cell attached to the mixing chamber remains in the
liquid state. The liquid pressures and temperatures, when the
plug is formed and when the \heyon cools down to the liquid/solid
coexistence point, are monitored by a pressure sensor at the
mixing chamber. The change in pressure within the cell itself is
sensed from the oscillation amplitude of one of the modes during
solidification.  After the plug is formed, the total time elapsed
to freeze \heyon in the cell was varied from several minutes to
four hours.

% defining superfluid density and dissipation
When our torsional oscillator with the sample cell filled with
solid \heyon is cooled, the two resonant frequencies ($f_i(T)$)
increase below 300 mK and reach maxima near 35 mK.  This increase
in frequency has been interpreted to arise from a partial
decoupling of solid \heyon from the sample cell wall and has been
called NCRI \cite{Leggett70,Kim05}.  The NCRI fraction (NCRIf) is
evaluated from our measurements as $\left[ f_i(T) - (f_{i0}(T) -
\Delta f_i)\right]/\Delta f_i$. The temperature dependent changes
in resonant frequencies are accompanied by decrease in quality
factor. A measure of dissipation induced by the presence of solid
\heyon is evaluated as $\delta\left[1/Q_i(T)\right] = 1/Q_i(T) -
1/Q_{i0}(T)$.  The data of resonant frequency and quality factor
reported here are obtained after stabilizing the temperature and
waiting for transient effects to diminish to negligible levels.

% description of data on NCRI and dissipation
In all, we have grown and studied five solid samples with final
solid pressures between 27 and 42 bar.  The observed features
were qualitatively similar in all solid samples.  Among the
samples we studied, we present results, as an example, of a 37
bar sample solidified over four hours.  The temperature of this
sample was maintained below 310 mK throughout all of the
measurements reported here after the initial cool down so as to
avoid any changes in the properties of our sample by some
annealing process.  The measured NCRIf and the dissipation while
keeping the rim velocity sufficiently low to avoid critical
velocity effects (see below) are shown in Fig. \ref{rhos}.  It
should first be noted that the maximum NCRIf observed is only 0.1
\%, which is considerably smaller than those found generally in
annular cells \cite{Kim04b,Rittner06}.  (The maximum NCRIf in
another sample solidified deliberately rapidly over several
minutes was larger at 0.12 \%.)  On the other hand, the maximum
is not so different from some of those found in cylindrical cells
in other laboratories \cite{Clark07,Shirahama07,Penzev07}.  At
$T$ $<$ 35 mK, NCRIf is independent of both temperature and
frequency. The dissipation is small in the same temperature
range.  In the higher temperature range $T$ $>$ 35 mK, both the
NCRIf and the dissipation show significant dependence on
frequency. The NCRIf measured with the first mode is lower than
that with the second mode where $T$ $>$ 35 mK.  At $T$ = 100 mK,
NCRIf($f_2$)/NCRIf($f_1$) = 1.9 and this ratio continues to
increase at higher temperatures.  The observed magnitude of
$\delta Q_i^{-1}$ is also smaller than in other annular cells. In
the temperature range, 35 mK $<$ $T$ $<$ 100 mK, $\delta
Q_1^{-1}$ is greater than $\delta Q_2^{-1}$ contrary to what
might be expected from the relative difference of NCRIf measured
with the two modes.  The dissipation maximum value of the first
mode is greater than that of the second by a factor of 1.7.  The
temperature where the dissipation maximum occurs is slightly
($\sim$ 4 mK) lower in the first mode than the second. The
temperature "width" of the dissipation peak is narrower in the
first mode.
\begin{figure}%[htbp]
\includegraphics[width=3in]{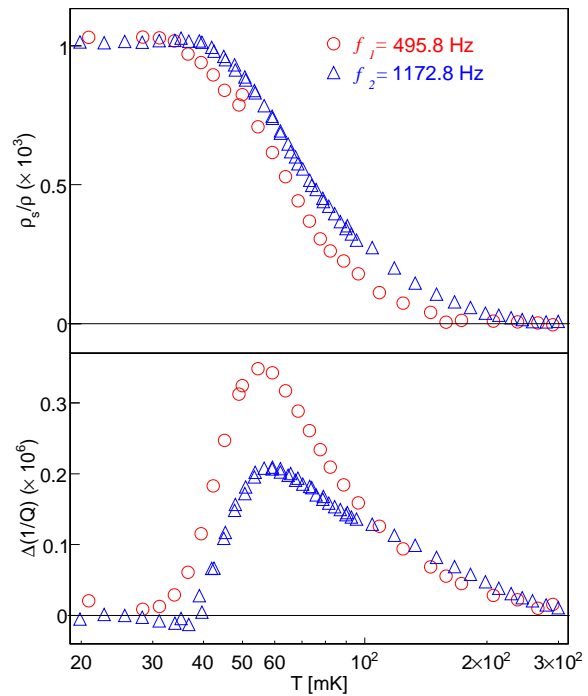}
\caption{\label{rhos} Temperature dependence of non-classical
rotational inertia fraction and change of dissipation (see text)
at $f_1$ ($\bigcirc$) and $f_2$ ($\triangle$). Pressure of the
solid is 37 bar. Rim velocities are 8.5 $\sim$ 13.5 $\mu$m/sec at
$f_1$, and 7.2 $\sim$ 16.0 $\mu$m/sec at $f_2$.}
\end{figure}

\begin{figure}[htbp]
\includegraphics[width=3in]{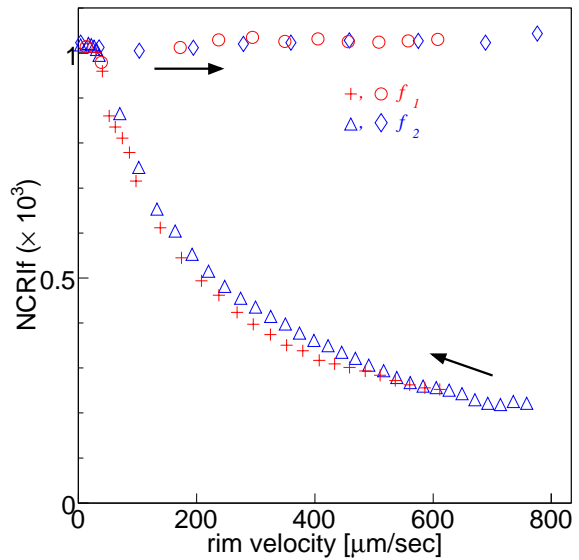}
\caption{\label{velocity_dep_19mK} Rim velocity dependence of
non-classical rotational inertia fraction for two frequencies at
the temperature of 19 mK. Measurements are taken as the drive is
decreased (+) (increased ($\bigcirc$)) in sequence from the
maximum (minimum) value shown for $f_1$. Similarly, measurements
are taken as the drive is decreased ($\triangle$) (increased
($\diamond$)) in sequence from the maximum (minimum) value shown
for $f_2$. Arrows indicate the direction of the rim velocity
changes.}
\end{figure}

% critical velocity effects and history dependence at low temperature
All previous torsional oscillator experiments have shown that the
NCRIf decreases as the drive level is increased and the decrease
has been ascribed as a "critical velocity" effect. It has not
been clarified if the reduction in NCRIf was controlled instead
by the amplitude of displacement or acceleration. To investigate
the critical velocity effect, the oscillator is cooled down to 19
mK in the first mode with a low velocity near 10 $\mu$m/s.  When
the drive level is increased, to our surprise, the NCRIf does not
diminish but remains constant up to 600 $\mu$m/s as shown by
circles in Fig. \ref{velocity_dep_19mK}. It is conceivable that
the NCRIf decays if we wait long enough, but the time constant
for this is estimated (assuming exponential decay for this
purpose) to be more than 10$^2$ hours.  Following the advice by
A. Clark and M. Chan \cite{Clark07}, the torsional oscillation is
initiated in the first mode with a relatively high drive level at
300 mK. While keeping the drive level constant, the oscillator is
then cooled down to 19 mK, where the rim velocity becomes 610
$\mu$m/s (shown by crosses). In this process, as the drive level
is decreased, the NCRIf increases and eventually attains the low
velocity limit consistent with Fig. \ref{rhos}.  This increase in
NCRIf as the rim velocity is decreased is similar to those
velocity effects as observed by others
\cite{Kim04a,Kim04b,Kim06,Shirahama07}.  If the oscillation is
initiated at any other rim velocity $<$ 610 $\mu$m/s at 300 mK
and cooled down to 19 mK, the observed NCRIf is that shown by the
cross in Fig. \ref{velocity_dep_19mK} at the corresponding rim
velocity. When similar procedures are followed with the second
mode, the same NCRIf is obtained when plotted as a function of
rim velocity (shown by triangles and diamonds).  Cooling the
oscillator at a low or high rim velocity is analogous to "zero
field cooling" or "field cooling" process, as suggested to us by
A. Clark and M. Chan \cite{Clark07}, in the studies of
superconducting and magnetic materials.

The reduction in NCRIf as the drive level is increased is
observed to be the same for both modes when it is plotted against
the cell rim velocity.  If NCRIf is plotted against displacement
amplitude or acceleration, the reduction does not coincide in the
two modes.  This observation demonstrates convincingly that it is
the rim velocity, not displacement amplitude (strain) nor
acceleration (stress) applied to the solid, that best
parameterize the reduction in NCRIf. The observed history
dependence of NCRIf on the initial state set by the rim velocity
at \emph{low} temperature is an important characteristics of the
supersolid state.

\begin{figure}[htbp]
\includegraphics[width=3in]{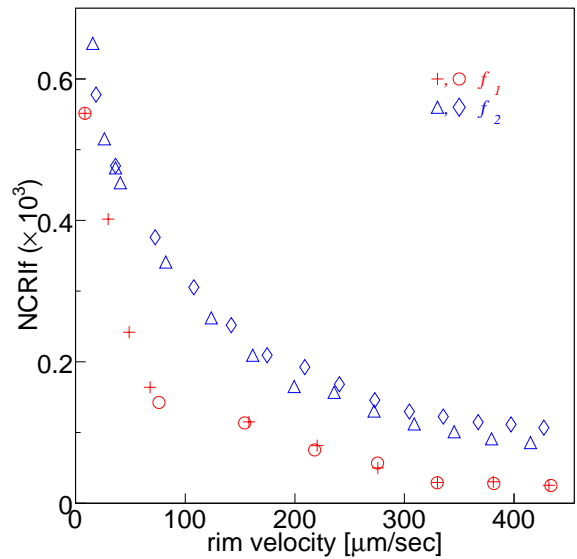}
\caption{\label{velocity_dep_62.5mK} Velocity dependence of
non-classical rotational inertia fraction for two frequencies at
63 mK. Symbols refer to the same processes as in Fig.
\ref{velocity_dep_19mK}.}
\end{figure}

% critical velocity and reversibility at high temperature
Our measurements of critical velocity effects at 63 mK are shown
in Fig. \ref{velocity_dep_62.5mK}.  The circles (crosses) and
diamonds (triangles) show NCRIf when the rim velocity is
increased (decreased) for the first and second mode,
respectively. The NCRIf decreases as the rim velocity is
increased.  When the difference at this temperature at low
velocity (see Fig. \ref{rhos}) is added to the NCRIf($f_1$), the
reduction is best parameterized by the rim velocity just as at 19
mK.  The saturation of the NCRIf below 15 $\mu$m/s as seen in
Fig. \ref{velocity_dep_19mK} is absent. There is no history
dependence at 63 mK!  The measured NCRIf does not depend on how
the initial state is reached. The border between the history
dependent low temperature behavior and the reversible higher
temperature behavior appears to be around 40 mK.  This is close
to the temperature where the NCRIf begins to decrease.

% discussion on vortex liquid theory by Anderson
Anderson interprets the observed NCRI as a consequence of vortex
liquid and supercurrents flowing around a thermally excited
fluctuation of vortices  \cite{Anderson07}.  He suggests that there
is a relaxation rate at which vortices can move in and out of the
sample. The rate is thought to decrease as the temperature is
lowered and as the \hesan impurity concentration is increased.
\hesan is thought to act promote pinning.  When the torsional
oscillator frequency matches the rate of vortex motion, a maximum in
dissipation would be observed. According to the model, the
dissipation peak should occur at higher temperature for higher
oscillator frequency.  Our data shows that the dissipation peak for
the second mode occurs at a slightly higher (by 4 $\pm$ 1 mK)
temperature than that of the first mode.  The direction of change in
the peak temperature is consistent with the Anderson model, but a
quantitative comparison is not possible (see Fig. 1 of Ref.
\cite{Anderson07}).  If the Anderson model is correct, the vortex
flow rate is greater than 500 s$^{-1}$ at $T$ $\geq$ 50 mK.  It
would then be expected that the NCRI is reversible, as indeed seen
in our experiment, at the time scale (of order 10$^2$ s) of our
measurements. At low temperatures, $T$ $\leq$ 35 mK, however, the
observed relaxation rate is dependent on the history of velocity
variation as illustrated in Fig. \ref{velocity_dep_19mK}.  The
vortex motion rate is apparently dependent strongly on the existing
number of vortices.

From the point of view of vortex liquid model, the irreversible
increase in NCRIf as seen in Fig. \ref{velocity_dep_19mK} from the
high to low rim velocity may be interpreted as a process in which
vortices can easily escape presumably from the cell surface.  Once
the vortices are absent, or nearly absent, they are apparently
excluded from entering solid \heyon even when the rim velocity is
increased.  These observations are reminiscent of Meissner effect in
superconductors.  A theoretical discussion of vortex nucleation in
solid \heyon has been given by Saslow \cite{Saslow05}.

% discussion on Huse model
According to the phenomenological two component model developed
by Huse and Khandker \cite{Huse07},  if a solid sample is
homogeneous, the change ($\delta f$) in the torsional oscillator
frequency in the $T$ = 0 limit is related to the maximum damping
at the apparent supersolid transition as $\delta f/f \sim \delta
(1/Q)_{max}$. Evaluating the quantities from the data shown in
Fig. \ref{rhos}, we find the ratio $(\delta f/f)/\delta
(1/Q)_{max}$ to be 2.7 and 6.5 for the first and second mode,
respectively.  The ratios are comparable to those of the "best"
among the samples studied by Kim and Chan \cite{Kim06}. One might
expect that the ratio should be independent of frequency if it is
simply a measure of homogeneity.  The temperature width of the
dissipation peak is also thought to indicate the degree of sample
inhomogeneity.  It is apparent in Fig. \ref{rhos} that the second
mode shows a broader temperature width in $\delta (1/Q)$ than the
first.

% discussion about glassy solid He-4 (Nussinov et al.)
Nussinov \textit{et al.} \cite{Nussinov06} proposed that the
observed increase in frequency and the peak in dissipation of
torsional oscillators containing solid \heyon might be explained
in terms of solidification process of liquid-like component into
a glass at low temperature.  In their model of the glass state,
there is a characteristic relaxation time $s$ which increases as
temperature is decreased.  When the inverse relaxation time
matches the oscillator frequency, a peak in the oscillator
dissipation appears.  The change in frequency-dependent
dissipation in their model naturally leads to a monotonic
increase in frequency as temperature is decreased. If $s$ is
assumed to take the simple form \cite{Nussinov06}, $s=s_0
e^{\Delta/k_BT}$, where $s_0$ and $\Delta$ are parameters, the
measured dissipations and frequency shifts of the two modes
cannot be reconciled.  Other forms of $s$ might reduce
discrepancy.  Our observations place severe constraints on the
glass model to explain the origin of observed NCRI effects of
solid $^4$He.
% discussion about defect vibration model by Iwasa
% summary paragraph
%In summary, the non-classical rotational inertia of a cylindrical
%solid \heyon measured at 1172 Hz differs from that at 497 Hz by a
%factor varying from unity at temperatures below 35 mK to 2 near
%100 mK.  The dissipation as measured by $Q^{-1}$ is small for
%both modes at temperatures below 35 mK, and it becomes near 60 mK
%a maximum which is greater by a factor 1.7 for the \textit{lower}
%than the higher mode.  These frequency dependent effects are
%difficult to reconcile with mechanical models proposed to account
%for the non-classical rotational inertia.  Irreversible and
%history dependent effects tend to support interpretation in terms
%of vortices which move or become pinned depending on the process.

% acknowledgment
We thank A.C. Clark and M.H.W. Chan for discussing their
experimental results and the relevance to vortex motion prior to
publication.  We are grateful to W.M. Saslow, P.W. Anderson, W.F.
Brinkman, D.A. Huse, J.D. Reppy, K. Shirahama and Z. Nussinov for
stimulating discussions. Y. A. thanks the Japan Society for
Promotion of Science for providing postdoctoral fellowship and
JCG Aresty Foundation for support of undergraduate research. This
work has been supported by the National Science Foundation
through grant DMR-0456862.

\bibliographystyle{apsrev}
\bibliography{supersolid,ballistic_phonon}

\end{document}